\documentstyle[prl,aps,epsf]{revtex}
\draft
\begin{document}
\twocolumn[\hsize\textwidth\columnwidth\hsize\csname @twocolumnfalse\endcsname

\title{Fibrillar templates and soft phases in systems with short-range
dipolar and long-range interactions}
\author{C.J. Olson Reichhardt, C. Reichhardt, and A.R. Bishop}
\address{
Theoretical Division,
Los Alamos National Laboratory, Los Alamos, New Mexico 87545}

\date{\today}
\maketitle
\begin{abstract}
We analyze the thermal fluctuations
of 
particles 
that have a short-range dipolar
attraction and a long-range 
repulsion.  
In an 
inhomogeneous particle density
region, 
or ``soft phase,''
filamentary patterns appear
which are destroyed only at very high temperatures.
The filaments 
act as 
a fluctuating template for correlated percolation 
in which
low-energy excitations can move through the stable pattern by 
local rearrangements.
At intermediate temperatures, 
dynamically averaged checkerboard states appear.
We discuss possible implications for cuprate superconducting and related
materials.
\end{abstract}
\vspace{-0.1in}
\pacs{PACS numbers: 73.50.-h, 71.10.Hf}
\vspace{-0.3in}

\vskip2pc]
\narrowtext
Mesoscopic and inhomogeneous
ordering of charges in diverse materials such as 
high-T$_c$ superconductors \cite{Emery93},
colossal magnetoresistant manganites \cite{Uehara,Burgy},
and two-dimensional (2D) electron gas systems \cite{Fradkin}
is a subject of intense recent study, as such ordered
phases may be highly relevant for understanding 
the properties of these materials, including
superconductivity mechanisms, pseudo gaps, 
transport responses, and magnetic responses.
Such phases include labyrinths, stripes, and clusters, and 
often consist of intermediately ordered  
pattens, lying between completely ordered 
and completely disordered systems \cite{PhillipsPRL}.
Inhomogeneous charge ordering phases can be produced by a competition
between repulsive and attractive interactions. 
In the case of the metal oxides, 
holes with a repulsive Coulomb interaction move in an antiferromagnetic
background, with the distortion of the 
spins giving rise to a dipolar attraction between holes, and allowing
the formation of clump, Wigner crystal, and stripe phases.
A variety of other microscopic mechanisms that produce 
a similar coexistence of 
long-range repulsion competing with directional short range 
attraction occur in a wide range of systems,
such as defects or dislocations
in elastic media, covalent glasses \cite{Phillips},
and systems with a finite density of 
Jahn-Teller polarons \cite{Mertelj}. 
Systems with 
effective competing repulsive-attractive interactions are abundant in 
soft matter, where bubble and stripe phases are observed \cite{Andelman}. 
A particularly relevant issue in 
these inhomogeneous systems is the role of temperature.
Open questions include
how the mesoscopically ordered patterns fluctuate or melt, and 
what the relevant dynamical modes are. 
Another important issue is what measures 
best characterize
these systems.  
Temperature effects are particularly important since dynamical and 
morphological changes along filaments
couple to changes in the material properties of the sample.  

In this work we focus on 
the thermal properties of a quasiclassical model 
for charge ordering of holes in transition
metal oxides, in which the particles have a Coulomb repulsion and 
a dipolar attraction. As a function of hole density (doping) we observe
an extended region that we term a ``soft phase'' 
comprised of partially ordered
filaments.  Ordered clumps form for densities below this region, and
ordered stripes (Wigner crystal-like phases)
occur above it.  We find that
the soft filamentary structures persist 
to high temperatures.  Within the soft phase
region there is a low temperature onset of motion along the filaments 
coupled with fluctuations of the filamentary structures.
The filaments act as a template for {\it correlated percolation} of 
particle motion. 
When the particle positions are averaged over long times,
the filaments form an ordered checkerboard pattern.    
We have also considered a simpler model of long-range repulsion and
{\it isotropic} short range attraction, and again find evidence for a 
filamentary soft phase, suggesting that the presence of soft phases
with low temperature correlated percolation may be a generic feature of 
systems with competing long/short range interactions.    
However, the filamentary soft phase is amplified by the anisotropy
(e.g. dipolar in this present case), and the dynamics is typically 
softer on filaments than in clusters.

Our phenomenological model for the holes consists of particles with
competing long-range and short-range interactions \cite{prl03epl}.
Four charge-ordering phases arise depending on the hole
density and the strength of a dipolar interaction induced by the holes:
a Wigner crystal; a diagonal stripe phase;
an intermediate geometric phase; and at low hole densities and larger
dipole interaction strengths, a clump phase.  The transitions among these
four phases have been studied in simulations \cite{BrankoPRL,BrankoPRB}.
We tune the doping level in our model by directly varying the hole
density in a rectangular computational box of size
$L_x \times L_y$, with $L_x$ and $L_y$ up to 100 unit cells in a
CuO$_2$ plane.  We initialize the simulation by placing the holes at
random and assigning each hole a magnetic dipole moment of fixed size but
random 
direction.  We use an efficient Monte Carlo method to find the
minimum of the total potential in this 
system \cite{BrankoPRB}.

We base our model for the interaction between the charges on the spin
density wave picture of the transition metal oxides.  Full details
of the model can be found in \cite{BrankoPRL,BrankoPRB}.  
The effective interaction between two holes, 1 and 2, a distance
${\bf r}$ apart, where $r$ is relaxed to an arbitrary 
(continuous, off-lattice) value,
is given by

\begin{figure}
\center{
\epsfxsize=3.5in
\epsfbox{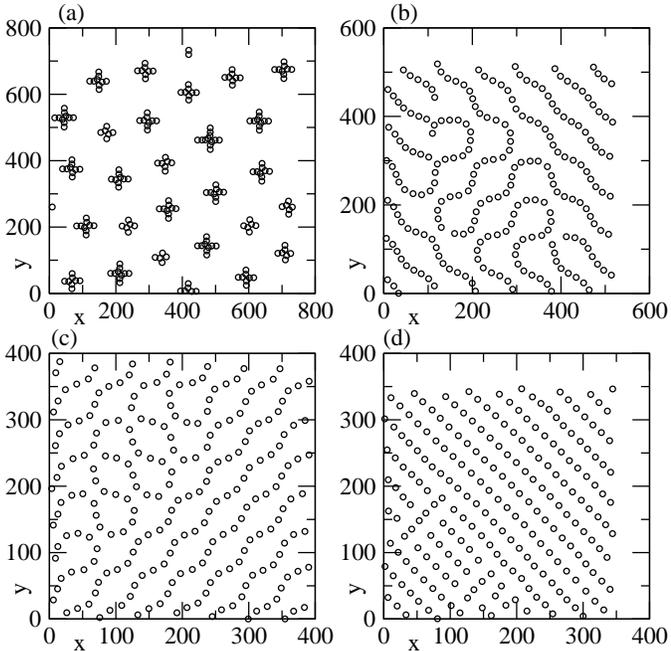}}
\caption{Static positions of holes for different densities.
(a) Clump phase, $n=0.6$, (b) soft phase, $n=1.2$, 
(c) soft phase, $n=2.1$, (d) anisotropic Wigner crystal phase,
$n=2.7$.
}
\label{fig:static}
\end{figure}

\noindent
\begin{equation}
V({\bf \tilde{r}})=
\frac{q^2}{\tilde{r}} - Ae^{-\tilde{r}/a}-
B\cos(2\theta-\phi_1-\phi_2)e^{-\tilde{r}/\xi}.
\end{equation}
Here, $q=1$ is the hole charge, $\theta$ is the angle between ${\bf r}$ and
a fixed axis, and $\phi_{1,2}$ are the angles of the magnetic dipoles
relative to the same fixed axis, which we assume can take an arbitrary
value.  
$A$ is the strength of the short-range anisotropic interaction,
and $B$ 
is that of the magnetic dipolar interaction 
[$B \approx A/(2\pi\xi^2)$], which we assume to be independent
variables.  Throughout this work we take $A=0$.  For the magnetic
correlation length, we assume the approximate dependence
$\xi=3.8/\sqrt{n}$ \AA; however, in order to observe a larger
number of clusters for a fixed number of holes $N=225$, we artificially
reduce the value of $\xi$ to a smaller value $\xi^{\prime}$, with
 $\xi^{\prime}=\sqrt{0.15}\xi$.
Test simulations on much larger systems of 2000 particles with the full
screening length indicate that the physics is qualitatively
unchanged by this screening length reduction.
The distance $\tilde{r}$ is measured in units of
the cuprate lattice spacing $a_0=3.8$ \AA, and
the hole density is given by $n$. 
The sample is periodic in the $x-y$ plane \cite{jensen}.

We first consider the 
static hole 
configurations, illustrated in Fig.~\ref{fig:static}, obtained at different
hole densities by annealing the system from a high temperature.  
At
low densities, cross-like clumps form and organize into 
an ordered lattice structure [Fig.~\ref{fig:static}(a)].
Above a density of 
$n=0.9$ the clumps begin to touch and are replaced by 
partially disordered filamentary patterns of the 
type illustrated in Fig.~\ref{fig:static}(b) for
$n=1.2$ and Fig.~\ref{fig:static}(c) for $n=2.1$. 
These 
filamentary  patterns persist within a glassy window 
that extends up to $n=2.4$,when 
a more ordered

\begin{figure}
\center{
\epsfxsize=3.5in
\epsfbox{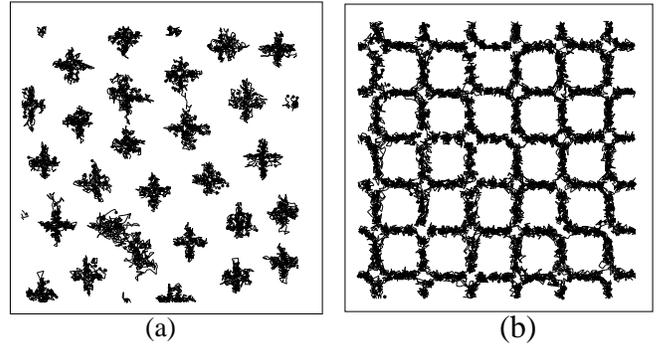}}
\caption{Image of system above the filamentary melting transition, with
lines indicating the motion of the holes between consecutive
simulation frames.
(a) $n=0.6$ in the clump phase, where the particles remain
predominantly localized. 
(b) $n=1.2$ showing the square modulated liquid phase in the
filamentary regime. 
}
\label{fig:square}
\end{figure}

\noindent
anisotropic Wigner crystal pattern forms, as shown in Fig.~\ref{fig:static}(d)
for $n=2.7$.


The transition from the filamentary pattern phase of Fig.~\ref{fig:static}(b,c)
to the anisotropic Wigner crystal phase of Fig.~\ref{fig:static}(d) occurs
when the interactions between the holes become dominated by the Coulomb
repulsion.  If the system formed a perfect isotropic Wigner
crystal, the lattice constant of this crystal would be 
$r_{WC}=L_x/\sqrt{N}$.  
We can 
estimate the density at which the 
transition occurs by a balance of forces, 
$q^2/\tilde{r}_{WC}^2 \approx B/\xi$.  This becomes $a_0^2/B=r_{WC}$,
giving $n=\sqrt{10 B/a_0}$, or $n=2.3$ for the parameters considered here,
in reasonable agreement 
with the transition 
observed in the simulations.

We next consider the role of temperature-induced 
fluctuations of these patterns.
For the clump region seen in Fig.~\ref{fig:static}(a), 
the superlattice  clump 
structure remains stable
up to a very high temperature. The  superlattice and
clumps break up simultaneously  
around $T_m=930$ K, where the melting temperature $T_m$ is 
measured by the onset of 
diffusion (see inset to Fig. 4). 
In general particles are confined within the clump up to this
temperature, although occasionally a particle can escape from the edge of
clump and move to another clump.
Such limited motion of particles within  the 
clumps begins at around 700K. 
The anisotropic Wigner
crystal phases found at densities $n>2.4$ beyond the filamentary soft phase 
are stable up to about
700K.  Above this temperature 
the entire pattern melts rapidly, although 
we have not determined the order of 
this transition.              

We now consider the fluctuations within the soft phase.
Well below $T_m$, we find a modulated liquid phase, in which
the charges are constrained to remain within a 
filamentary square pattern, but are free to move along the
filaments.  This is illustrated in Fig.~\ref{fig:square}(b) for
$n=1.2$ in the soft phase, showing the holes moving along
a square pattern.
The pattern is oriented with the
simulation box due to the residual dipoles, which form because we
have not restricted the spin on each charge to take discrete 

\begin{figure}
\center{
\epsfxsize=3.5in
\epsfbox{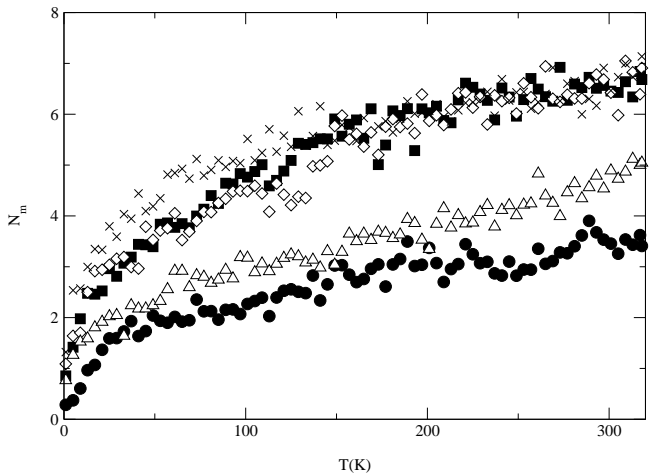}}
\caption{Pattern reorientation measure $N_m$ as a function of temperature
for different hole densities: filled circles, $n=0.75$, filled squares,
$n=1.2$, open diamonds, $n=1.65$, x's, $n=2.25$, and open triangles,
$n=2.7$.}
\label{fig:nm}
\end{figure}

\noindent
values
but have instead allowed the angles to vary continuously.
The dipoles align along $\pm 45^\circ$, and thus the system is
divided into two equal but interconnected basins of magnetization.
We contrast the filamentary square pattern observed throughout the
soft phase with a similar 
image of the clump phase,
shown in Fig.~\ref{fig:square}(a) for
$n=0.6$.  Here the holes
remain localized within each clump. 

The modulated liquid phase remains stable up to the temperature
$T_m$ at which the pattern itself is destroyed.
We can define a second, lower melting temperature $T_s$ at which
the onset of the 
square (checkerboard) fluid phase, or modulated liquid, occurs.  
In this modulated liquid phase the particle motion is constrained
to follow the square template.  
Although the hole density is not high enough to create this
square structure in the static configuration, when the holes are moving
their effective density increases, allowing them to form the square
state. At
any moment the square state is not fully formed; in a
snapshot of the system, the square state contains imperfections and
density modulations.  On average over time, however, the square state
is present, and the particles move in response to its structure.

The temperature $T_s$ at which the square modulated 
liquid appears
varies with doping.  More disordered hole configurations (such as
$n=1.2$)
have a lower $T_s$
than more ordered configurations (such as 
$n=2.7$).
To quantify this, we 
must detect the transition to
the modulated liquid.  The onset of the modulation occurs when
the structure of the pattern begins to change globally due to local
rearrangements of the holes.  For example, for 
$n=1.2$,
one of the
loops may open and reconnect into a new loop formed out of a
previously straight segment.  We can detect changes in the
structure of the pattern by monitoring the local orientation 
$\Theta_v$ (in the range $[0,\pi/2)$) of
a Voronoi polygon centered on each particle 

\begin{figure}
\center{
\epsfxsize=3.5in
\epsfbox{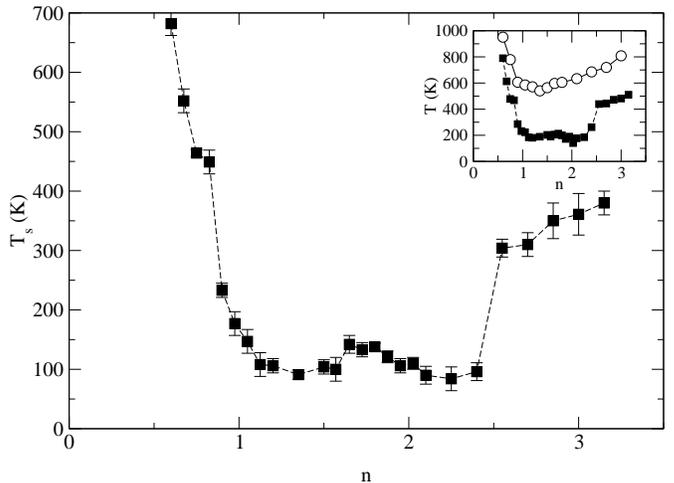}}
\caption{Onset temperature $T_s$ of modulated square liquid state as
a function of hole density $n$. Inset: Melting temperature 
of pattern $T_m$
(open circles),
and $T_s$ (black squares).}
\label{fig:tmelt}
\end{figure}

\noindent
at fixed $T$.
Changes in the pattern orientation produce
increased fluctuations of the local orientation $<\sqrt{\Theta_v^2}>$.
We find the number of particles $N_m$ undergoing
pattern changes by counting the particles that have
$<\sqrt{\Theta_v^2}> \ge 0.2\pi$ 
during a fixed
number of MC steps.
The onset of the modulated liquid should occur when enough particles
are fluctuating on average to allow the excitations to percolate through
the system.  If the holes are spaced $\xi^{\prime}$ from
each other along a line, $L_x/\xi^{\prime}=38$ holes fit across
the sample length,  
in $N\xi^{\prime}/L_x=5.9$ rows of closely packed particles.  We therefore
select $N_m = 6$ as the filamentary transition, since this would
allow one fluctuation at each crossing point of the square pattern to
extend across the system.  The temperatures $T_s$
obtained with this value of $N_m$ agree well with the onset of the
modulated liquid phase as observed through animations of the simulations.
We find a clear decrease of $T_s$ in the soft phase when the
modulated liquid appears, as 
illustrated in Fig.~\ref{fig:nm}.  

We plot $T_s$ as a function of density in Fig.~\ref{fig:tmelt},
as determined from measurements of $N_m$.
We find a dramatic decrease in $T_s$ over a soft window from
$n=0.9$ to $n=2.4$.
This corresponds to the geometrically disordered percolation
phase.  Outside of this soft window, particle motion does not occur
until significantly higher temperatures are reached, especially 
for $n<0.9$.
Using different values of $N_m$ to
determine $T_s$ does not alter the range of densities at which the soft
window appears; it merely shifts the curve slightly in
temperature.

In the soft window, we observe {\it correlated} percolation.  
{\it The onset
of structural percolation coincides with a highly directional
softness.}  This is in contrast to the isotropic melting that occurs at
much higher temperatures.  Thus, we have a system which contains a 
{\it rigid}
template of {\it soft} filaments.  
At $T_s$, a type of dislocation mobility transition occurs in which
a bundle excitation becomes delocalized.
In this case, because the effective interaction between
particles along the chain is highly nonlinear, 
propagation
of a soliton-like excitation can occur, resulting in the time-averaged
square pattern.

In the model considered here, the short-range attractive interaction
between particles has a strong directional dependence.
We have also studied a system with isotropic short
range attractive interactions \cite{physicaD}, 
and 
find that in the stripe phase, melting first 
occurs along the length of each stripe.  
We have measured the melting temperature $T_m$ of the three phases 
in the isotropic model by 
computing the particle diffusion at each temperature. 
Particles diffuse freely within
the stripes but do not enter the regions between the stripes until higher
temperatures.  In this system the stripe structure is not 
filamentary, so only 
constrained two-dimensional motion of the particles within
the broad stripe cluster
is possible.  
$T_m$ is lower
in the stripe phase; however, we do not find as dramatic a decrease in 
$T_m$ due to the quasi-two-dimensional nature
of the particle motions within the stripe.

We now discuss some implications of our results for experiments
on metal-oxide materials. Our results indicate that the charge-ordered
states should persist up to very high temperatures. 
Signatures of disordered filamentary states occur at much lower
temperatures with a transition to a square (checkerboard) 
state at intermediate 
temperatures.
However, the short/long-range interactions only appear upon (polaron-like)
localization of holes, which onsets at the ``pseudogap temperature.''
Above this temperature, a more metallic electronic state is
expected.  
The square phase that we observe is a time-averaged, not a static,
structure, so we predict that it would be detected in experiments such as
STM which resolve time-averaged images, 
and it may be a good template for
the STM images observed in doped cuprates 
\cite{Davis,Kapitulnik}.
Short-time imaging techniques
may reveal only disordered structures.
The low temperature phase at $T<T_s$ is not fluctuating strongly,
and thus appears more disordered than the square phase at $T>T_s$, which 
could lead to an observable disorder-order ``transition'' with
increasing $T$.
Noise measurements could also detect the onset
of the square modulated liquid phase: in the clump phase
there should be little noise, while in the modulated liquid phase, large
scale fluctuations of the patterns should produce increased
noise.  
It is also likely that external fields will easily
induce currents along the filamentary paths in the soft phase.
The soft phase at $T_s<T<T_m$ also shows
similarities with the inhomogenous states observed in manganites
between the true critical temperature $T_c$ and a higher temperature
$T^*$ at which short-range order first appears \cite{Burgy}.

Some stripe-based theories for superconductivity 
require fluctuating stripes 
\cite{Stripes}. 
In our system, we find that the soft 
regions (which appear in an intermediate doping region)
dynamically fluctuate at low $T$.
Our soft phase could then 
correspond to the fluctuating stripe 
regime.  An important feature of the soft phase we observe is that 
the fluctuations are predominantly on
percolating filaments rather than meandering of the filaments
themselves.
The fluctuating checkerboard state may thus
provide a good theoretical
starting point for introducing quantum mechanical effects.

In summary, 
we have considered a model 2D system with competing long-range 
and directional short-range interactions. We find a soft filamentary
or fibrillar regime 
that occurs at densities between a low-density clump phase and a 
high-density anisotropic crystalline phase.
This soft phase consists of partially ordered
filamentary patterns of charges. 
At low temperatures, the filamentary patterns act as a rigid template
for a correlated percolation of particles along the pattern: within
the soft phase, the onset of motion occurs at a far lower temperature
than that for melting the template backbone. (Quantum-mechanical
mesoscopic tunneling of the template can occur but at much slower
rates.)
We have also found a similar soft phase for other models with 
competing long/short
range interactions, indicating that these phases may be general features
of such systems.

We acknowledge helpful discussions with J.C. Phillips and A.B. Saxena.
This work was supported by the US DoE under Contract No. 
W-7405-ENG-36.

\end{document}